\pdfoutput=1

\documentclass[11pt,twoside,a4paper,cmspaper,final,collab]{cms-tdr}
\def\svnVersion{373366}\def\cmsCernNoTag{CERN-EP/2016-174}\def\cmsCernDate{\today}\def\cmsMessage{Published in Physics Letters B as \href{http://dx.doi.org/10.1016/j.physletb.2016.10.054}{\doi{10.1016/j.physletb.2016.10.054.}}}

\begin{document}\cmsNoteHeader{SMP-16-001}

\hyphenation{had-ron-i-za-tion}
\hyphenation{cal-or-i-me-ter}
\hyphenation{de-vices}
\RCS$Revision: 365291 $
\RCS$HeadURL: svn+ssh://svn.cern.ch/reps/tdr2/papers/SMP-16-001/trunk/SMP-16-001.tex $
\RCS$Id: SMP-16-001.tex 365291 2016-08-23 18:28:12Z alverson $
\newlength\cmsFigWidth
\ifthenelse{\boolean{cms@external}}{\setlength\cmsFigWidth{0.48\textwidth}}{\setlength\cmsFigWidth{0.75\textwidth}}
\ifthenelse{\boolean{cms@external}}{\providecommand{\cmsLeft}{top\xspace}}{\providecommand{\cmsLeft}{left\xspace}}
\ifthenelse{\boolean{cms@external}}{\providecommand{\cmsRight}{bottom\xspace}}{\providecommand{\cmsRight}{right\xspace}}
\providecommand{\ZZ}{\ensuremath{\cPZ\cPZ}\xspace}
\providecommand{\pp}{\ensuremath{\Pp\Pp}\xspace}
\newcommand{\elfour}{\ensuremath{\ell^+\ell^-\ell^{\prime+}\ell^{\prime-}}\xspace}
\newcommand{\elfourtau}{\ensuremath{\ell^+\ell^-\ell^{\prime\prime+}\ell^{\prime\prime-}}\xspace}

\cmsNoteHeader{SMP-16-001}
\title{Measurement of the ZZ production cross section and $\cPZ \to \elfour$ branching fraction in pp collisions at $\sqrt{s} = 13\TeV$}

\author{CMS Collaboration}

\date{\today}

\abstract{
Four-lepton production in proton-proton collisions,
${\Pp\Pp} \to \left(\cPZ / \gamma^*\right)\left(\cPZ/ \gamma^*\right) \to
\elfour$,
where $\ell, \ell' = \Pe$ or $\PGm$, is studied at
a center-of-mass energy of 13\TeV with the CMS detector at the LHC. The data
sample corresponds to an integrated luminosity of 2.6\fbinv.
The \ZZ production cross section,
$\sigma(\Pp\Pp \to \cPZ\cPZ) = 14.6 ^{+1.9}_{-1.8}\stat
^{+0.5}_{-0.3}\syst \pm 0.2\thy  \pm 0.4\lum\unit{pb}$, is measured
for events with two opposite-sign, same-flavor lepton pairs
produced in the mass region
$60 < m_{\ell^+\ell^-}, m_{\ell^{\prime +}\ell^{\prime -}} < 120\GeV$.
The Z boson branching fraction to four leptons is measured to be
$\mathcal{B}(\cPZ \to \elfour) = 4.9 _{-0.7}^{+0.8}\stat
_{-0.2}^{+0.3}\syst _{-0.1}^{+0.2}\thy \pm 0.1\lum
\times 10^{-6}$ for
the four-lepton invariant mass in the range
$80 < m_{\elfour} < 100\GeV$
and dilepton mass $m_{\ell^+\ell^-} > 4\GeV$ for
all opposite-sign, same-flavor lepton pairs. The results are in agreement with
standard model predictions.
}

\hypersetup{%
pdfauthor={CMS Collaboration},%
pdftitle={Measurement of the ZZ production cross section and Z to l+l-l'+l'- branching fraction in pp collisions at sqrt(s) = 13 TeV},%
pdfsubject={CMS},%
pdfkeywords={CMS, physics, electroweak}}

\maketitle

\section{Introduction}

Measurements of diboson production at the CERN LHC allow precision
studies of the standard model (SM).
These measurements are important for testing predictions that were
recently made available at next-to-next-to-leading-order (NNLO) in
quantum chromodynamics (QCD)~\cite{Cascioli:2014yka}. Comparing these predictions
to data at a range of center-of-mass energies gives insight into the structure
of the electroweak gauge sector of the SM, and new proton-proton collision
data at $\sqrt{s} = 13\TeV$ allow diboson measurements at the highest energies
to date. Any deviations from expected values could be an indication of
physics beyond the SM.

Previous measurements of the ZZ production cross section from CMS were performed in
the $\cPZ\cPZ \to \elfourtau$ and $\cPZ\cPZ \to \ell^+\ell^-\nu\nu$
decay channels, where $\ell = \Pe, \Pgm$ and
$\ell^{\prime\prime} = \Pe, \Pgm, \Pgt$ for both Z bosons produced on-shell, in the dilepton
mass range 60--120\GeV~\cite{Chatrchyan:2012sga, CMS:2014xja, Khachatryan:2015pba}.
These measurements were made with data sets corresponding to
integrated luminosities of 5.1\fbinv at $\sqrt{s} = 7\TeV$ and
19.6\fbinv at $\sqrt{s} = 8\TeV$, and
agree with SM predictions.
The ATLAS Collaboration produced similar
results at $\sqrt{s} = 7$, 8, and 13\TeV~\cite{Aad:2012awa, Aad:2015rka, Aad:2015zqe},
which also agree with the SM.

Extending the mass window for the dilepton candidates to lower values
allows measurements of $\left(\cPZ/\gamma^{\ast}\right) \left(\cPZ/\gamma^{\ast}\right)$ production,
where ``$\cPZ$'' may indicate an on-shell $\cPZ$ boson or an off-shell $\cPZ^{\ast}$ boson.
The resulting sample includes Higgs boson events in the ``golden channel''
$\PH  \to  \cPZ\cPZ^{\ast}  \to  \elfour$, where $\ell' = \Pe, \Pgm$, 
and rare $\cPZ$ boson decays to four leptons. The
$\cPZ  \to  \ell^+\ell^-\gamma^{\ast}  \to  \elfour$
decay was studied in detail at LEP~\cite{Buskulic:1994gk} and was
observed in pp collisions by CMS~\cite{CMS:2012bw} and by
ATLAS~\cite{Aad:2014wra}. Though the branching fraction for this decay is orders of
magnitude smaller than that for the $\cPZ \to \ell^+\ell^-$ decay, the
precisely known mass of the $\cPZ$ boson makes the four-lepton mode
useful for calibrating mass measurements of the nearby Higgs
resonance.

This letter reports a study of four-lepton production
($\pp  \to  \elfour$, where $\ell$ and $\ell'$ indicate
electrons or muons) at $\sqrt{s} = 13\TeV$ with a data set corresponding to
an integrated luminosity of $2.62 \pm 0.07$\fbinv recorded in 2015. From this study,
cross sections are inferred for nonresonant production of pairs of $\cPZ$ bosons,
$\pp  \to  \cPZ\cPZ$, where both $\cPZ$ bosons are produced on-shell,
defined as the mass range 60--120\GeV, and resonant $\pp  \to  \cPZ
\to \elfour$
production. Discussion of resonant Higgs boson
production is beyond the scope of this letter.

\section{The CMS detector}

A detailed description
of the CMS detector, together with a definition of the coordinate system used
and the relevant kinematic variables, can be found
in Ref.~\cite{Chatrchyan:2008zzk}.

The central feature of the CMS apparatus is a superconducting solenoid of
6\unit{m} internal diameter, providing a magnetic field of 3.8\unit{T}.
Within the solenoid volume are a silicon pixel and strip
tracker, a lead tungstate crystal electromagnetic calorimeter (ECAL), and a
brass and scintillator hadron calorimeter (HCAL),
which provide coverage in pseudorapidity $\abs{ \eta } < 1.479 $ in a barrel
 and $1.479 < \abs{ \eta } < 3.0$ in two endcap regions.
Forward calorimeters extend the
coverage provided by the barrel and
endcap detectors to $\abs {\eta} < 5.0$. Muons are measured in gas-ionization detectors embedded in
the steel flux-return yoke outside the solenoid
in the range $\abs{\eta} < 2.4$, with
detection planes made using three technologies: drift tubes, cathode strip
chambers, and resistive plate chambers.

Electron momenta are estimated by combining energy measurements in the
ECAL with momentum measurements in the tracker. The momentum resolution for
electrons with transverse momentum $\pt \approx 45\GeV$
from $\Z \to \Pep \Pem$ decays
ranges from 1.7\% for nonshowering electrons in the barrel region to 4.5\% for
showering electrons in the endcaps~\cite{Khachatryan:2015hwa}.
Matching muons to tracks measured in
the silicon tracker results in a $\pt$ resolution for
muons with $20 <\pt < 100\GeV$ of 1.3--2.0\% in the barrel and better than~6\%
in the endcaps. The \pt resolution in the barrel is better than 10\% for muons
with \pt up to 1\TeV~\cite{Chatrchyan:2012xi}.

\section{Signal and background simulation}
\label{sec:mc}

Signal events are generated with
\POWHEG~2.0~\cite{Alioli:2008gx,Nason:2004rx,Frixione:2007vw} at
next-to-leading-order (NLO) in QCD for quark-antiquark processes and
leading-order (LO) for quark-gluon processes. This includes
$\ZZ$, $\cPZ\gamma^\ast$, $\cPZ$, and $\gamma^\ast\gamma^\ast$ production
with a constraint of $m_{\ell^+\ell^{\prime -}} > 4\GeV$ applied between all pairs
of oppositely charged leptons at the generator level to avoid infrared divergences.
The $\Pg\Pg  \to  \ZZ$ process is simulated at LO
with \MCFM~v7.0~\cite{Campbell:2010ff}. These samples are scaled to
correspond to cross sections calculated at NNLO for
$\cPq\cPaq  \to  \ZZ$~\cite{Cascioli:2014yka} (scaling $K$ factor 1.1) and at NLO for
$\Pg\Pg  \to  \ZZ$~\cite{Caola:2015psa} ($K$ factor 1.7). The
$\Pg\Pg  \to  \ZZ$ process is calculated to
$\mathcal{O}\left(\alpha_s^5\right)$, where $\alpha_s$ is the strong coupling constant,
while the other contributing processes are calculated to
$\mathcal{O}\left(\alpha_s^4\right)$; this higher-order correction
is included because the effect is known to be large~\cite{Caola:2015psa}.

A sample of Higgs boson events is produced in the gluon-gluon fusion
process with \POWHEG~2.0
in the NLO QCD approximation. The Higgs boson
decay is modeled with
\textsc{jhugen}~3.1.8~\cite{Gao:2010qx,Bolognesi:2012,Anderson:2013afp}.
The $\cPq\cPaq  \to  \PW\cPZ$ process is generated with \POWHEG~2.0.

The \PYTHIA~v8.175~\cite{Sjostrand:2006za,Sjostrand:2015,Alioli:2010xd} package
is used for parton showering, hadronization, and
the underlying event simulation, with parameters set by the CUETP8M1
tune~\cite{Khachatryan:2015pea}. The NNPDF3.0 ~\cite{nnpdf}
set is used as the default set of parton distribution functions (PDFs).
For all simulated event samples, the PDFs are calculated to the same
order in QCD as the process in the sample. 

The detector response is simulated using a detailed
description of the CMS detector implemented with the \GEANTfour
package~\cite{GEANT}. The event reconstruction is performed with
the same algorithms used for data.
The simulated samples include additional interactions per bunch crossing,
referred to as ``pileup.''
The simulated events are weighted so that the pileup distribution matches
the data, with an average of about 11 interactions per bunch
crossing.

\section{Event reconstruction}
\label{sec:eventreconstruction}

All long-lived particles in each collision event --- electrons,
muons, photons, and charged and neutral hadrons --- are identified and reconstructed
with the CMS particle-flow (PF) algorithm~\cite{CMS-PAS-PFT-09-001, CMS-PAS-PFT-10-001}
from a combination of the signals from all subdetectors.
Reconstructed electrons~\cite{Khachatryan:2015hwa} and muons~\cite{Chatrchyan:2012xi}
are candidates for inclusion in four-lepton final states
if they have
$\pt^\Pe > 7\GeV$ and $\abs{\eta^\Pe} < 2.5$ or
$\pt^\Pgm > 5\GeV$ and $\abs{\eta^\Pgm} < 2.4$. These are designated
``signal leptons.''

Signal leptons are also required to originate from the event vertex, defined as the
proton-proton interaction vertex whose associated charged particles have the
highest sum of $\pt^2$.
The distance of closest approach between each lepton track and the event vertex
is required to be less than 0.5\unit{cm} in the plane transverse to the beam axis,
and less than 1\unit{cm} in the direction along the beam axis.
Furthermore, the significance of the three-dimensional impact parameter relative
to the event vertex, $\mathrm{SIP_{3D}}$, is required to satisfy
$\mathrm{SIP_{3D}} \equiv \abs{ \mathrm{IP} / \sigma_\mathrm{IP}} < 4$
for each lepton, where $\mathrm{IP}$ is the distance
of closest approach of each lepton track to the event vertex
and $\sigma_\mathrm{IP}$ is its associated uncertainty.

Signal leptons are required to be isolated from other particles in the event. The
relative isolation is defined as
\begin{equation}
        R_\text{iso} = \bigg[ \sum_{\substack{\text{charged} \\ \text{hadrons}}} \!\! \pt \, + \,
                             \max\big(0, \sum_{\substack{\text{neutral} \\ \text{hadrons}}} \!\! \pt
                                       \, + \, \sum_{\text{photons}} \!\! \pt \, - \, \pt^\mathrm{PU}
                                       \big)\bigg] \bigg/ \pt^{\ell},
        \label{eq:iso}
\end{equation}
where the sums run over the charged and neutral hadrons, and photons, in a
cone defined by
$\Delta R \equiv \sqrt{\smash[b]{\left(\Delta\eta\right)^2 + \left(\Delta\phi\right)^2}} < 0.3$
around the lepton trajectory, where $\phi$ is the azimuthal angle in radians.
To minimize the contribution of charged particles from pileup to the isolation calculation,
charged hadrons are included only if they originate from the
event vertex. The contribution of
neutral particles from pileup is $\pt^\mathrm{PU}$. For electrons, $\pt^\mathrm{PU}$
is evaluated with the ``jet area'' method described in Ref.~\cite{Cacciari:2007fd};
for muons, it is taken
to be half the sum of the $\pt$ of all charged particles in the cone originating
from pileup vertices. The factor
one-half accounts for the expected ratio of charged to neutral particle energy
in hadronic interactions. A lepton is considered isolated if
$R_\text{iso} < 0.35$.

Emission of final-state radiation (FSR) photons by the signal leptons may
degrade the performance of the isolation requirements and $\cPZ$ boson mass
reconstruction. These photons are omitted from the isolation
determination for signal leptons and are implicitly included in dilepton kinematic
calculations.
Photons are FSR candidates
if $\pt^\gamma > 2\GeV$, $\abs{\eta^\gamma} < 2.4$, their relative
isolation (defined as in Eq.~(\ref{eq:iso}) with $\pt^\mathrm{PU} = 0$)
is less than 1.8, and
$\Delta R \left(\ell, \gamma \right) < 0.5$ with respect to the nearest signal lepton.
To avoid double counting of bremmstrahlung photons that are already included in electron
reconstruction, photons are not FSR candidates if there is any signal electron within
$\Delta R \left(\gamma, \Pe \right) < 0.15$ or within
$\abs{ \Delta \phi \left(\gamma,\Pe \right)} < 2$ and
$\abs{ \Delta \eta \left(\gamma,\Pe \right)}< 0.05$.
Because FSR photons have a higher average energy than photons from pileup
and are expected to be mostly collinear with the emitting lepton,
a photon candidate is accepted as FSR if
$\Delta R \left( \ell, \gamma \right) / \left(\pt^\gamma\right)^2 < 0.012\GeV^{-2}$.

In simulated $\cPZ\cPZ \to \elfour$ events,
the efficiency to select generated FSR photons is around 55\%, and roughly 85\%
of selected photons are matched to FSR photons. At least one FSR photon is
identified in approximately 2\%, 5\%, and 8\% of simulated
events in the $4\Pe$, $2\Pe 2\Pgm$, and $4\Pgm$ channels, respectively.
In data events with two on-shell $\cPZ$ bosons, no FSR photons are selected in
the $4\Pe$ decay channel, while at least one FSR photon
is selected in three and five events in the $2\Pe2\Pgm$ and $4\Pgm$ decay
channels, respectively.

The lepton reconstruction, identification, and isolation efficiencies
are measured with a tag-and-probe
technique~\cite{CMS:2011aa} applied to a sample of $\cPZ  \to  \ell^+\ell^-$ data events.
The measurements are performed in several bins of $\PT^{\ell} $ and $ |\eta^\ell|$.
The electron reconstruction and selection efficiency in the ECAL barrel (endcaps) varies from
about 85\% (77\%) at $\PT^{\Pe} \approx 10\GeV$
to about 95\% (89\%) for $\PT^{\Pe} \geq 20\GeV$,
while in the barrel-endcap transition region this efficiency is about 85\% averaged 
over all electrons with $\pt^{\Pe} > 7\GeV$.
The muons are reconstructed and identified with efficiencies above ${\sim}98\%$
within $\abs{\eta^{\Pgm}} < 2.4$.

\section{Event selection}

The primary triggers for this analysis require the presence of a
pair of loosely isolated leptons of the same or different flavors.
The highest \pt lepton must have $\pt^\ell > 17\GeV$, and the
subleading lepton must have
$\pt^\Pe > 12\GeV$ if it is an electron or $\pt^\Pgm > 8\GeV$
if it is a muon. The dielectron and dimuon triggers require that the
tracks corresponding to the leptons originate from within
2~mm of each other in the plane transverse to the beam axis. Triggers
requiring a triplet of lower-\pt leptons
with no isolation criterion, or a single high-\pt electron without an
isolation requirement, are also used.
An event is used if it passes any trigger regardless of the decay channel.
The total trigger efficiency for events within the acceptance of this
analysis is greater than 98\%.

A signal event must contain at least two
$\cPZ/\gamma^{\ast}$ candidates, each formed from an oppositely charged
pair of isolated signal electrons or muons.
Among the four leptons, the highest \pt lepton must have $\pt > 20\GeV$, and
the second-highest \pt lepton must have $\pt^\Pe > 12\GeV$ if it is an electron
or $\pt^\Pgm > 10\GeV$ if it is a muon.
All leptons are required to be separated by
$\Delta R \left(\ell_1, \ell_2 \right) > 0.02$,
and electrons are required to be separated from muons by
$\Delta R \left(\Pe, \mu \right) > 0.05$.

Within each event, all
permutations of leptons giving a valid pair of $\cPZ/\gamma^{\ast}$
candidates are considered separately.
Within each $\elfour$ candidate, the
dilepton candidate with an invariant mass closest to 91.2\GeV, taken as
the nominal $\cPZ$ boson mass, is denoted $\cPZ_1$ and is required to have a
mass greater than 40\GeV. The other dilepton candidate is denoted $\cPZ_2$.
Both $m_{Z_1}$ and $m_{Z_2}$ are required to be less than 120\GeV.
All pairs of oppositely charged leptons in the candidate are required to
have $m_{\ell \ell'} > 4\GeV$ regardless of flavor.

If multiple $\elfour$
candidates within an event pass all selections,
the passing candidate with $m_{\cPZ_1}$ closest to
the nominal $\cPZ$ boson mass is chosen. In the rare case
of further ambiguity, which may arise in events with
five or more signal leptons, the $\cPZ_2$ candidate
that maximizes the scalar $\pt$ sum of the four leptons is chosen.

Additional requirements
are applied to select events for measurements of specific processes.
The $\pp  \to  \cPZ\cPZ$ cross section is measured using events where both
$m_{\cPZ_1}$ and $m_{\cPZ_2}$ are greater than 60\GeV.
The $\cPZ  \to \elfour$
branching fraction is measured using events with
$80 < m_{\elfour} < 100\GeV$, a range chosen
to retain most of the decays in the resonance while removing most other
processes with four-lepton final states.

\section{Background estimate}

The major background contributions arise from
$\cPZ$ boson and $\PW\cPZ$ diboson production in association with
jets and from \ttbar production.
In all these cases, particles from jet fragmentation satisfy both lepton identification and
isolation criteria, and are thus misidentified as signal leptons.

The probability for such objects to be selected
is measured from a sample of
$\cPZ + \ell_\text{candidate}$ events, where $\cPZ$ is a pair of
oppositely charged, same-flavor leptons that pass all analysis requirements and
satisfy $\abs{ m_{\ell^+\ell^-} - m_{\cPZ}} < 10\GeV$, where
$m_\cPZ$ is the nominal $\cPZ$ boson mass. Each event in this sample must have exactly one
additional object $\ell_\text{candidate}$ that passes relaxed identification requirements with
no isolation requirements applied. The misidentification probability for
each lepton flavor is defined as a ratio of the
number of candidates that pass the final
isolation and identification requirements to the total number in the sample,
measured in bins of lepton candidate $\pt$ and $\eta$.
The number of $\cPZ + \ell_\text{candidate}$ events is corrected for contamination
from WZ production, or ZZ production in which one lepton is not
reconstructed. These events have a third genuine, isolated lepton that must be excluded
from the misidentification probability calculation. The WZ contamination is suppressed by requiring the missing
transverse energy $\ETmiss$ to be below 25\GeV. The $\ETmiss$ is defined
as the magnitude of the missing transverse momentum vector $\ptvecmiss$,
the projection onto the plane transverse to the beams of the negative
vector sum of the momenta of all reconstructed particles in the event.
Additionally, the transverse mass
$m_{\mathrm{T}} \equiv \sqrt{\smash[b]{(\et^\ell + \ETmiss )^2 -
(\ptvec^\ell + \ptvecmiss )^2}}$
of $\ell_{\text{candidate}}$  and the missing
transverse momentum vector is required to be less than 30\GeV. The residual contribution
of $\PW\cPZ$ and $\cPZ\cPZ$ events, which may be up to a few percent of the
events with $\ell_\text{candidate}$ passing all selection criteria, is estimated
from simulation and subtracted.

To account for all sources of background events, two control samples are 
used to estimate the number of background events
in the signal regions. Both are defined to contain events with
a dilepton candidate satisfying all requirements ($\cPZ_1$) and
two additional lepton candidates $\ell^{\prime +}\ell^{\prime -}$.
In one control sample, enriched in $\PW\cPZ$ events, one 
$\ell^{\prime}$ candidate is required to satisfy the full
identification and isolation criteria and the other must fail the full criteria
and instead satisfy only relaxed ones; in the other, enriched in 
$\cPZ$+jets events, both $\ell^{\prime}$ 
candidates must satisfy the relaxed criteria, but fail the full criteria. 
The additional leptons must have opposite charge and the same
flavor ($\Pe^{\pm}\Pe^{\mp}, \Pgm^{\pm}\Pgm^{\mp}$).
From this set of events, the expected number of background events in the
signal region is obtained
by scaling the number of observed $\cPZ_1+\ell^{\prime +}\ell^{\prime -}$ events
by the misidentification probability for each lepton failing the selection.
Low-mass dileptons may be sufficiently collinear that their isolation
cones overlap, and their misidentification probabilities are therefore correlated.
To mitigate the effect of these correlations, only the
control sample in which both additional
leptons fail the full selection is used if
$\Delta R \left(\ell^{\prime +} , \ell^{\prime -} \right) < 0.6$.
The background contributions to the signal regions of
$\cPZ \to \elfour$ and
$\cPZ\cPZ \to \elfour$ are summarized in
Section~\ref{sec:xsec}.

\section{Systematic uncertainties}

Systematic uncertainties are summarized in Table~\ref{table:systematics}.
In both data and simulated event samples, trigger efficiencies are evaluated with
a tag-and-probe technique. The ratio between data and simulation is applied
to simulated events, and the size of the resulting change in expected yield is
taken as the uncertainty for the determination of the trigger efficiency.
This uncertainty is
around 2\% of the final estimated yield. For
$\cPZ  \to  \Pep\Pem\Pep\Pem$ events, the uncertainty increases to 4\%.

\begin{table}[htb]
\centering
\topcaption{
  The contributions of each source of signal systematic uncertainty in the
  cross section measurements. The integrated luminosity uncertainty and the PDF and scale
  uncertainties are considered separately. All other uncertainties are added
  in quadrature into a single systematic uncertainty. Uncertainties that vary by
  decay channel are listed as a range.
}
\begin{tabular}{lcc}
Uncertainty & $\cPZ  \to  4\ell$ & $\cPZ\cPZ  \to  4\ell$  \\
\hline
ID efficiency             & 2--6\%       & 0.4--0.9\% \\
Isolation efficiency      & 1--6\%       & 0.3--1.1\% \\
Trigger efficiency        & 2--4\%       & 2\%        \\
MC statistics             & 1--2\%       & 1\%        \\
Background                & 0.7--1.4\%   & 0.7--2\%   \\
Pileup                    & 0.4--0.8\%   & 0.2\%      \\
\hline
PDF                       & 1\%          & 1\%        \\
QCD Scales                & 1\%          & 1\%        \\
\hline
Integrated luminosity     & 2.7\%        & 2.7\%      \\
\end{tabular}
\label{table:systematics}
\end{table}

The lepton identification and isolation efficiencies in simulation are corrected
with scaling factors derived with a tag-and-probe method and applied as a
function of lepton $\pt$ and $\eta$.
To estimate the uncertainties associated with the tag-and-probe technique, the total yield is
recomputed with the scaling factors varied up and down by the tag-and-probe
fit uncertainties. The uncertainties associated with the identification efficiency in
the $\ZZ  \to  \elfour$
($\cPZ  \to  \elfour$)
signal regions are found to be 0.9\%\,(6\%) in the $4\Pe$ final state, 0.7\%\,(4\%)
in the  $2\Pe 2\mu$ final state, and 0.4\% (2\%) in the $4\mu$ final state.
The corresponding uncertainties associated with the isolation efficiency are
1.1\%\,(6\%) in the $4\Pe$ final state, 0.7\%\,(3\%) in the $2\Pe 2\mu$ final
state, and 0.3\%\,(1\%) in  the $4\mu$ final state. These uncertainties are
higher for $\cPZ  \to  \elfour$ events because the leptons
generally have lower \pt, and the samples used in the tag-and-probe method
have fewer events and more contamination from nonprompt leptons in this
low-\pt region.

Uncertainties due to the effect of factorization ($\mu_F$) and
renormalization ($\mu_R$) scale choice on the $ZZ \to \elfour$
acceptance are evaluated with \POWHEG and \MCFM by varying the scales up
and down by a factor of two with respect to the default values
$\mu_F = \mu_R = m_{\cPZ\cPZ}$. These variations are much smaller than 1\%
and are neglected. Parametric  uncertainties
($\mathrm{PDF} {+} \alpha_s$)
are evaluated using
the CT10~\cite{Lai:2010vv} and NNPDF3.0 sets and are found to be less than 1\%. The largest
difference between predictions from \POWHEG
and \MCFM with different scales and PDF sets, 1.5\%, is considered to be the
theoretical uncertainty in the acceptance calculation.
An additional theoretical uncertainty arises from scaling the \POWHEG
$\cPq\cPaq  \to  \ZZ$ simulated sample from its NLO cross section to the
NNLO prediction, and the \MCFM $\Pg\Pg  \to  \ZZ$ samples
from their LO cross sections to the NLO predictions. The change in the
acceptance corresponding to this scaling procedure is found to be 1.1\%.
All theoretical uncertainties are added in quadrature.

The largest uncertainty in the estimated background yield arises from
differences in sample composition between the $\cPZ + \ell$ control sample
used to calculate the lepton misidentification probability and the
$\cPZ + \ell^+\ell^-$ control sample. A further uncertainty arises
from the limited number of events in the $\cPZ + \ell$ sample. A
systematic uncertainty of 40\% of the estimated background yield is applied
to cover both effects. The size of this uncertainty varies by channel, but
is less than 1\% of the total expected yield.

The uncertainty in the integrated luminosity of the data sample
is 2.7\%~\cite{CMS-PAS-LUM-15-001}.

\section{Cross section measurements}
\label{sec:xsec}

\begin{figure}[htbp]
\centering
\includegraphics[width=0.48\textwidth]{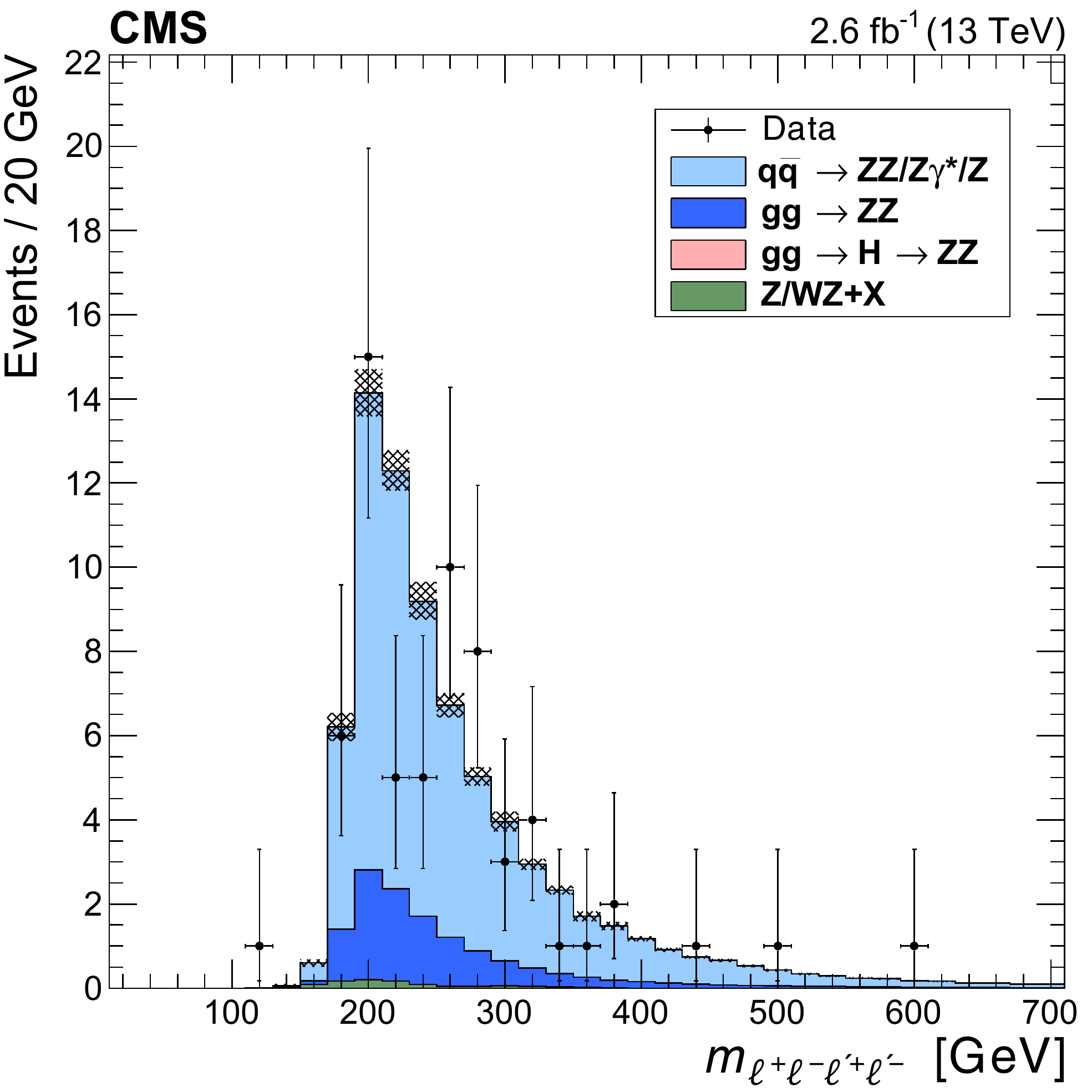}
\includegraphics[width=0.48\textwidth]{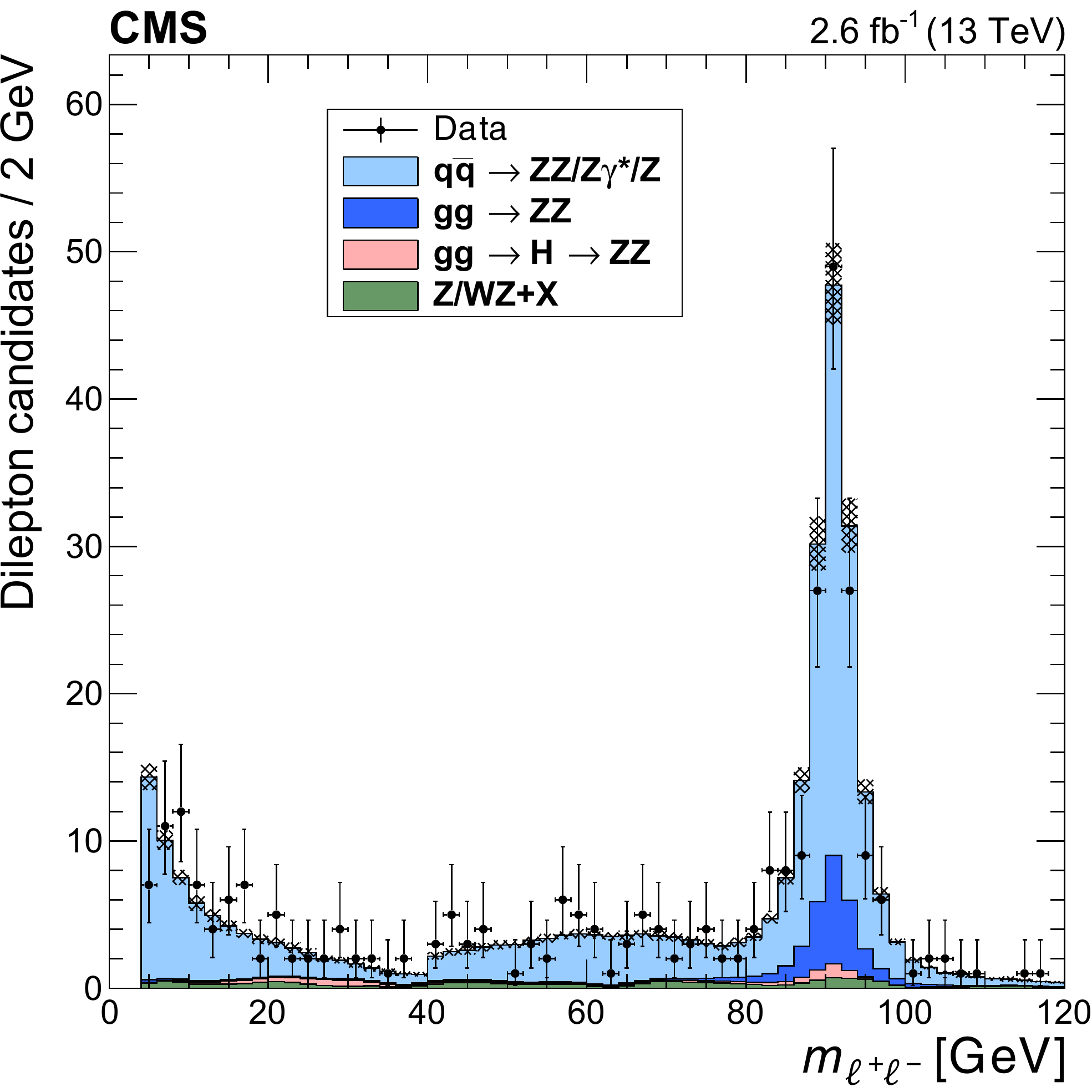}
\caption{
Distributions of (\cmsLeft) the four-lepton invariant mass $m_{\elfour}$
and (\cmsRight) the invariant mass of the dilepton candidates in all selected
four-lepton events, including both $\cPZ_1$ and $\cPZ_2$ in each event.
Points represent the data, while shaded histograms represent
the SM prediction and background estimate.
Hatched regions around the predicted yield
represent combined statistical, systematic, theoretical, and integrated
luminosity uncertainties. }
\label{fig:results_full}
\end{figure}

\begin{figure}[htbp]
\centering
\includegraphics[width=0.48\textwidth]{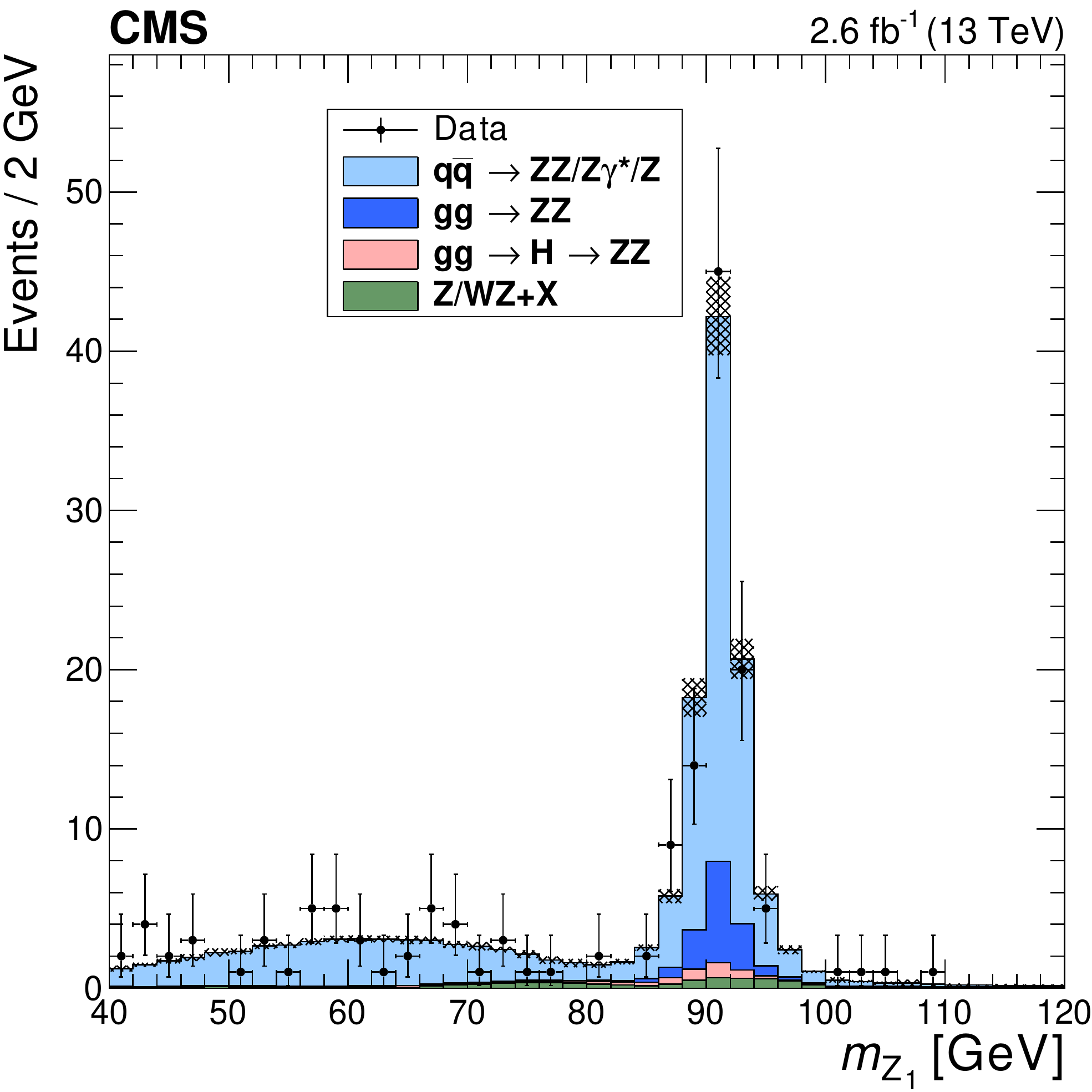}
\includegraphics[width=0.48\textwidth]{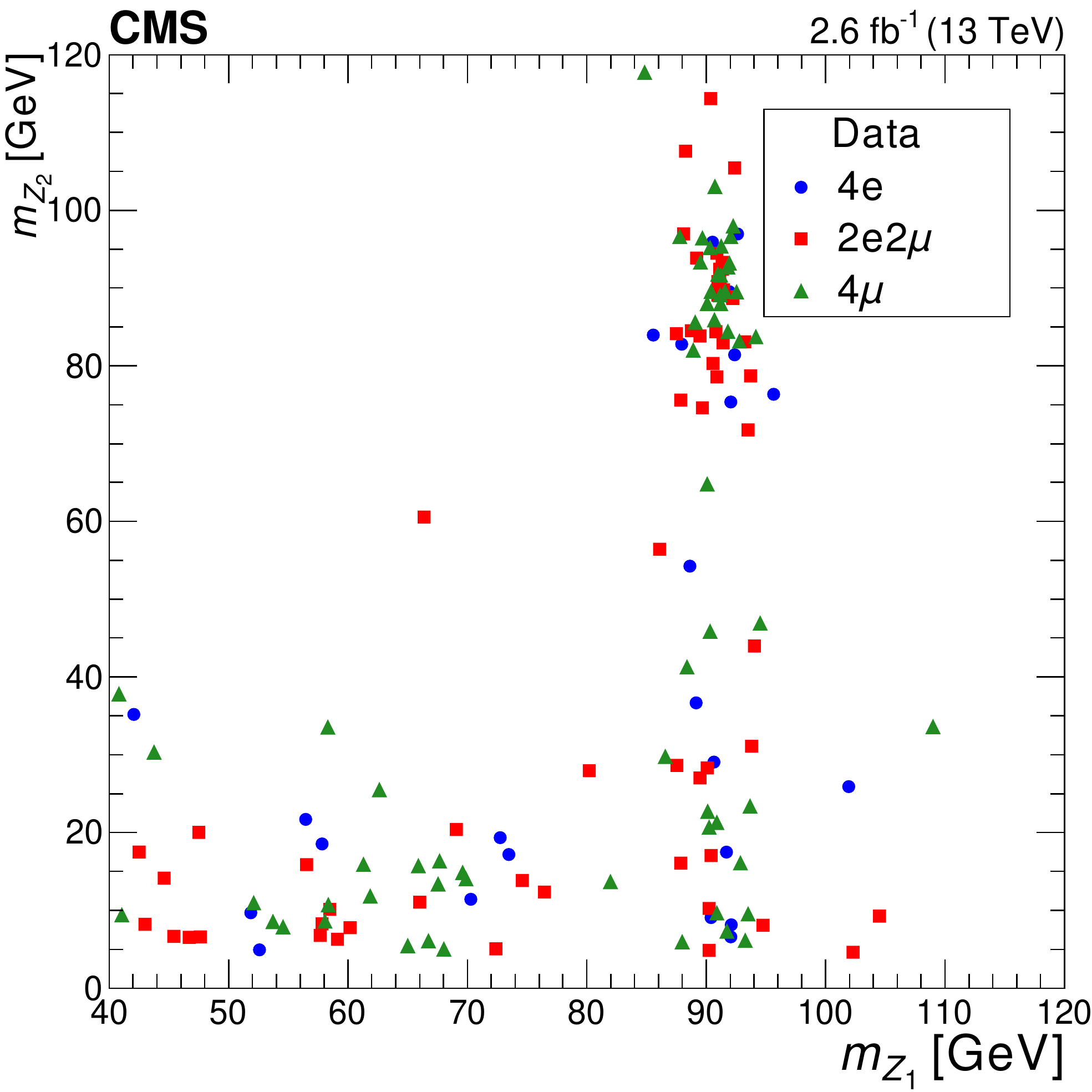}
\caption{
(\cmsLeft) The distribution of the reconstructed mass of the $\cPZ_1$ candidate.
Points represent the data, while shaded histograms represent
the SM prediction and background estimate.
Hatched regions around the predicted yield
represent combined statistical, systematic, theoretical, and integrated
luminosity uncertainties. (\cmsRight) The reconstructed $m_{\cPZ_2}$ plotted
against the reconstructed $m_{\cPZ_1}$ in data events, with
distinctive markers for each final state. }
\label{fig:results_full_Z}
\end{figure}

The distributions of the four-lepton mass and the masses of
the $\cPZ_1$ and $\cPZ_2$ candidates are shown in Fig.~\ref{fig:results_full}.
The SM predictions include nonresonant $\cPZ\cPZ$ predictions normalized
using the NNLO cross section, production of the SM Higgs
boson with mass 125\GeV~\cite{Aad:2015zhl}, and
resonant $\cPZ \to \elfour$ production.
The background estimated from
data is also shown.
The reconstructed invariant mass of the $\cPZ_1$ candidates, and a scatter plot
showing the correlation between $m_{\cPZ_2}$ and $m_{\cPZ_1}$ in data events,
are shown in Fig.~\ref{fig:results_full_Z}. In the scatter plot, 
clusters of events corresponding to $\cPZ\cPZ \to \elfour$,
$\cPZ\gamma^\ast \to \elfour$, and $\cPZ \to \elfour$ production can be seen.

\begin{figure}[htbp]
\centering
\includegraphics[width=0.48\textwidth]{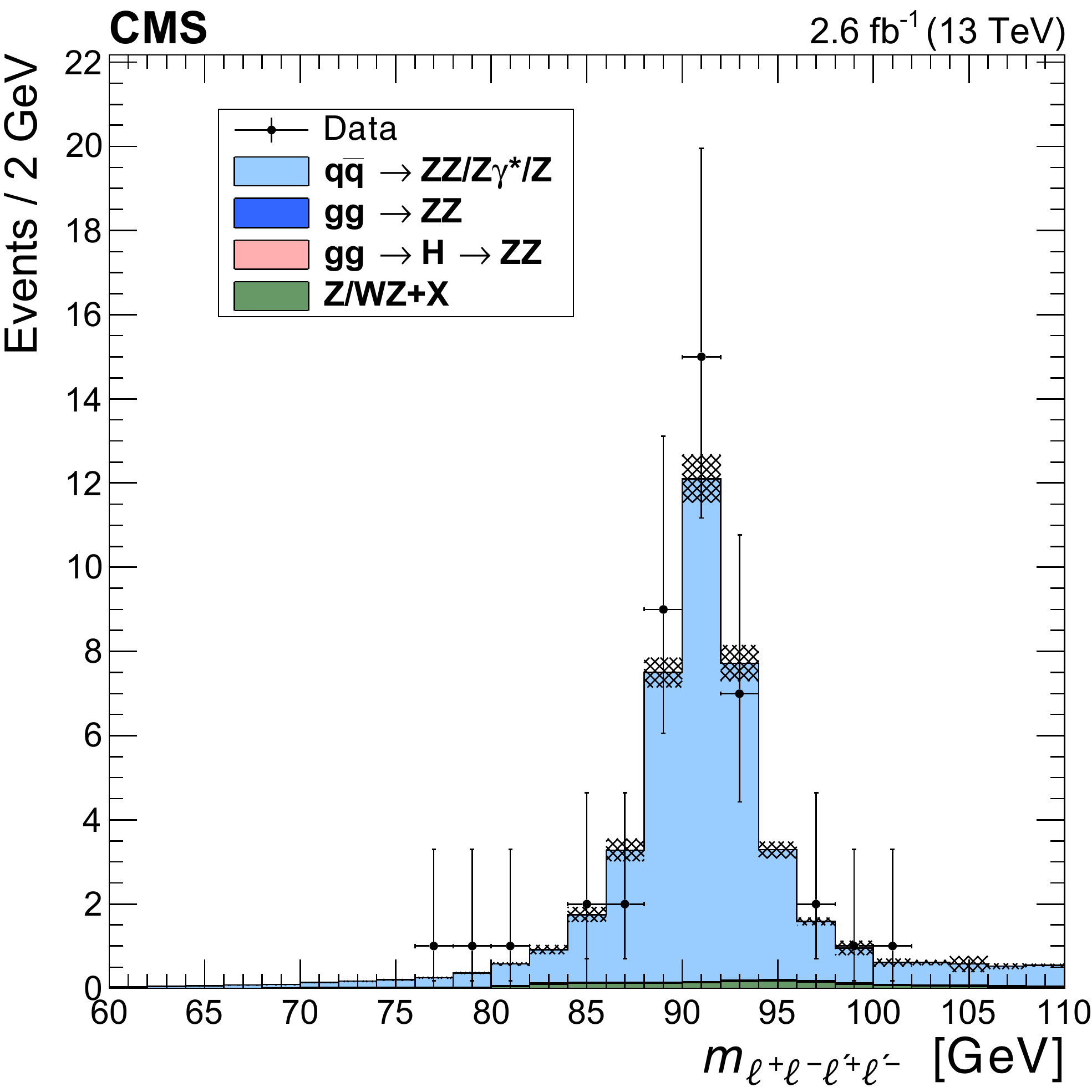}
\includegraphics[width=0.48\textwidth]{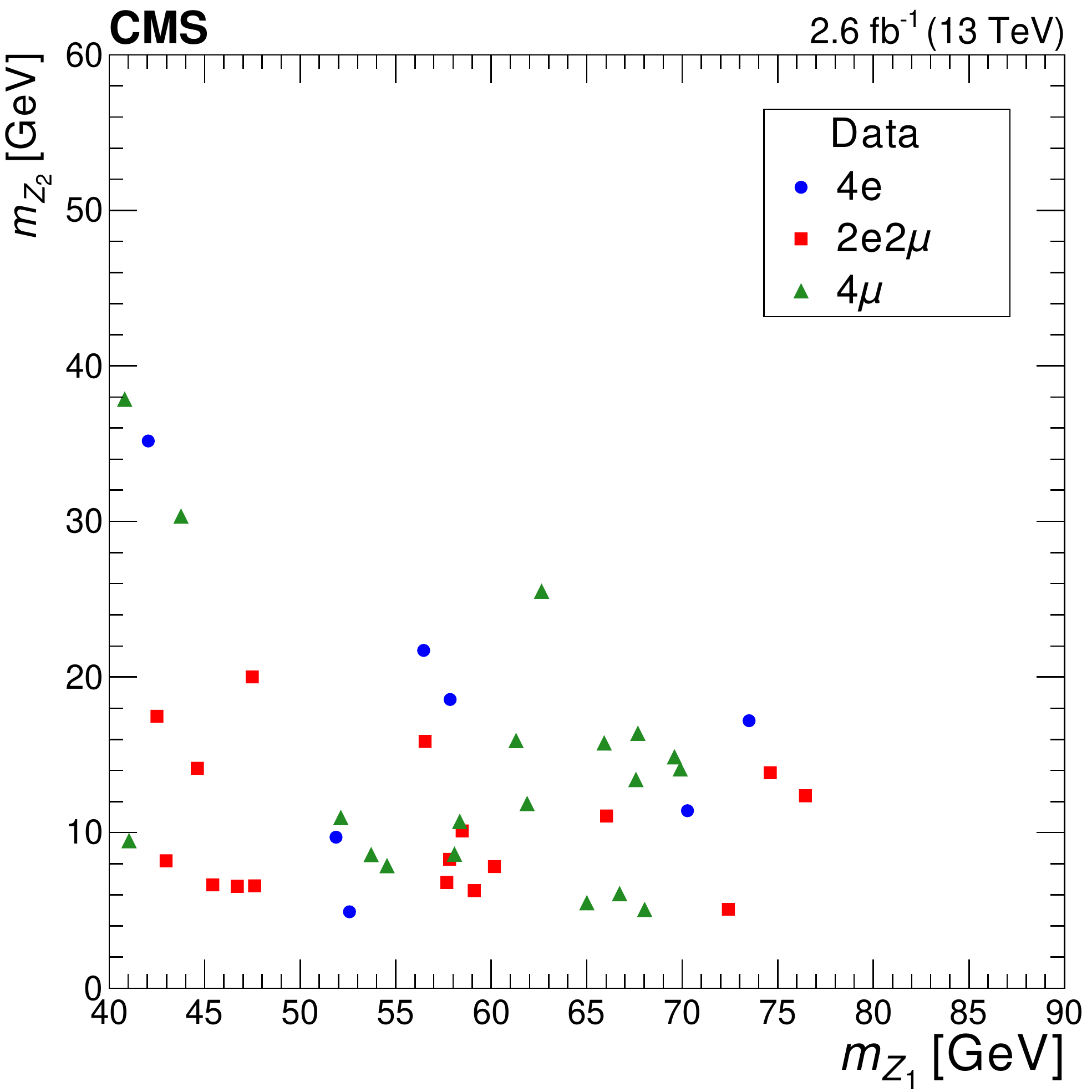}
\caption{
(\cmsLeft) The distribution of the reconstructed four-lepton mass
$m_{\elfour}$ for events selected with $m_{\elfour} < 110\GeV$.
Points represent the data, while shaded histograms represent
the SM prediction and background estimate.
Hatched regions around the predicted yield
represent combined statistical, systematic, theoretical, and integrated
luminosity uncertainties. (\cmsRight) The reconstructed $m_{\cPZ_2}$ plotted
against the reconstructed $m_{\cPZ_1}$ in data events selected with
$m_{\elfour}$ between 80 and 100\GeV, with
distinctive markers for each final state. }
\label{fig:results_z4l}
\end{figure}

The four-lepton invariant mass distribution below 110\GeV
is shown in Fig.~\ref{fig:results_z4l} (\cmsLeft).
Figure~\ref{fig:results_z4l} (\cmsRight) shows $m_{\cPZ_2}$ plotted against
$m_{\cPZ_1}$ for events with
$m_{\elfour}$ between $80$ and 100\GeV,
and the observed and expected event yields
in this mass region are given in Table~\ref{table:results_z4l}.

\begin{table*}[htbp]
\centering
\topcaption{ The observed and expected yields of four-lepton
 events in the mass region
$80 < m_{\elfour} < 100\GeV$
and estimated yields of background
events evaluated from data, shown for each final state and
summed in the total expected yield.
The first uncertainty is statistical, the second one is systematic.
}
\begin{tabular}{ccccc}
Final & Expected &  Background   & Total & Observed \\
state & $N_{\elfour}$ &  & expected  & \\[0.2ex]
\hline
$4\Pgm$       & $ 16.88 \pm 0.14 \pm 0.62 $ & $ 0.31 \pm 0.30 \pm 0.12 $ & $ 17.19 \pm 0.33 \pm 0.63  $ & $ 17 $\\
$2\Pe 2\Pgm$  & $ 15.88 \pm 0.14 \pm 0.87 $ & $ 0.37 \pm 0.27 \pm 0.15 $ & $ 16.25 \pm 0.31 \pm 0.88  $ & $ 16 $\\
$4\Pe$        & $ 5.58  \pm 0.08 \pm 0.53 $ & $ 0.21 \pm 0.10 \pm 0.08 $ & $ 5.78  \pm 0.13 \pm 0.53  $ & $ 6  $\\
\hline
Total         & $ 38.33 \pm 0.21 \pm 1.19 $ & $ 0.89 \pm 0.42 \pm 0.22 $ & $ 39.22 \pm 0.47 \pm 1.21  $ & $ 39 $\\
\end{tabular}
\label{table:results_z4l}
\end{table*}

\begin{figure}[htbp]
\centering
\includegraphics[width=0.48\textwidth]{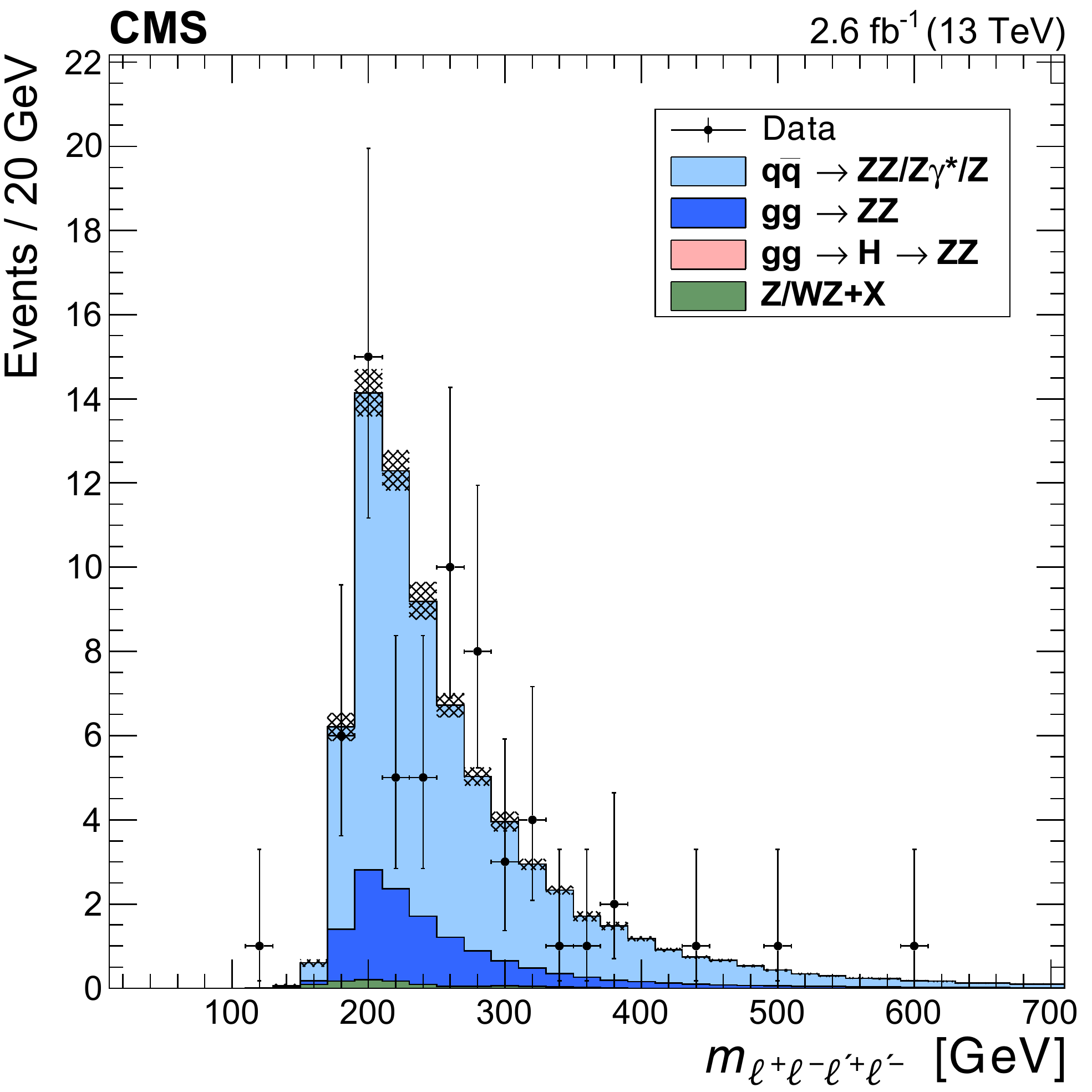}
\includegraphics[width=0.48\textwidth]{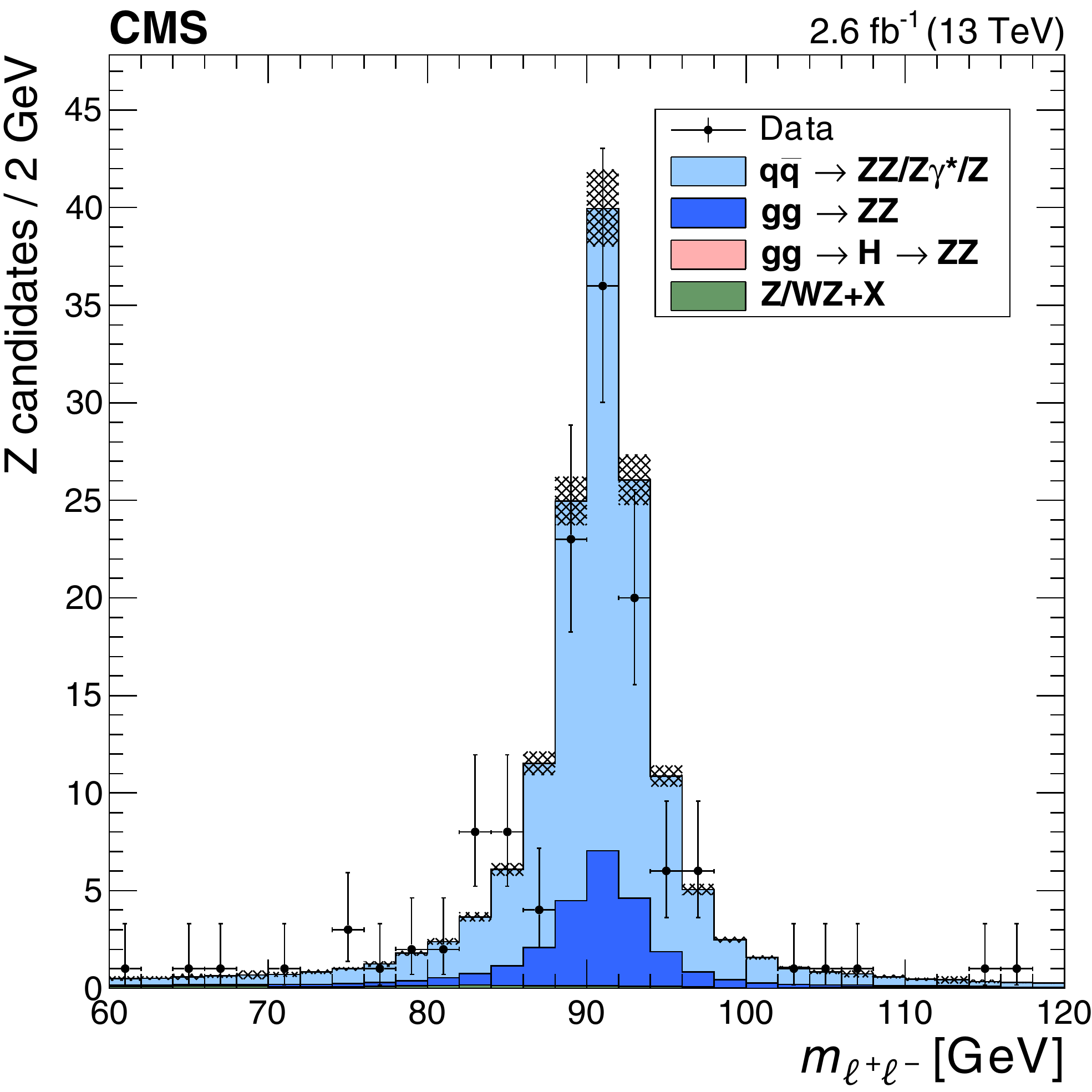}
\caption{
Distributions of (\cmsLeft) the four-lepton invariant mass $m_{\elfour}$
and (\cmsRight) dilepton candidate mass for four-lepton events selected with
both $\cPZ$ bosons on-shell.
Points represent the data, while shaded histograms represent
the SM prediction and background estimate.
Hatched regions around the predicted yield
represent combined statistical, systematic, theoretical, and integrated
luminosity uncertainties. }
\label{fig:results_smp}
\end{figure}

The reconstructed four-lepton invariant mass is shown  in Fig.~\ref{fig:results_smp} (\cmsLeft)
for events with two on-shell $\cPZ$ bosons.
Figure~\ref{fig:results_smp} (\cmsRight) shows the invariant mass distribution for all
$\cPZ$ candidates in these events.  The corresponding observed and expected
yields are given in Table~\ref{table:results_smp}.

\begin{table*}[htbp]
\centering
\topcaption{ The observed and expected yields of $\cPZ\cPZ$ events,
and estimated yields of background
events evaluated from data, shown for each final state and
summed in the total expected yield. The first uncertainty
is statistical, the second one is systematic.
}
\begin{tabular}{lcccc}
Final & Expected  &  Background   & Total & Observed \\
state & $N_{\elfour}$ &  & expected  & \\[0.2ex]
\hline \\[-1.9ex]
$4\Pgm$        & $ 21.80 \pm 0.15 \pm 0.46 $ & $ 0.00 ^{+0.24}_{-0.00} {}^{+0.10}_{-0.00}  $ & $ 21.80 ^{+0.28}_{-0.15} {}^{+0.47}_{-0.46}  $ & $  26 $\\
$2\Pe 2\Pgm$   & $ 36.15 \pm 0.20 \pm 0.81 $ & $ 0.60 \pm 0.34      \pm 0.24      $ & $ 36.75 \pm 0.34 \pm 0.85  $ & $  30 $\\
$4\Pe$         & $ 14.87 \pm 0.12 \pm 0.36 $ & $ 0.81 \pm 0.26      \pm 0.33      $ & $ 15.68 \pm 0.26 \pm 0.48  $ & $  8  $\\
\hline \\[-1.9ex]
Total          & $ 72.82 \pm 0.27 \pm 1.00 $ & $ 1.42 ^{+0.49}_{-0.43} {}^{+0.42}_{-0.41}  $ & $ 74.23 ^{+0.56}_{-0.45} {}^{+1.08}_{-1.08}  $ & $  64 $\\
\end{tabular}
\label{table:results_smp}
\end{table*}

The observed yields are used to evaluate the
$\pp  \to  \cPZ  \to  \elfour$
and $\pp  \to  \ZZ  \to \elfour$
production cross sections
from a combined fit to the number of observed
events in all the final states. The likelihood is a combination of
individual channel likelihoods for the signal and background hypotheses
with the statistical and systematic uncertainties in the form of scaling
nuisance parameters.
The ratio of the measured cross section to the SM cross section given by this fit including all channels
is scaled by the cross section used in the simulation to find the measured
fiducial cross section.

The definitions for the fiducial phase spaces for the
$\cPZ \to \elfour$ and
$\ZZ \to \elfour$
cross section measurements are given in Table~\ref{table:fiducial_cuts}.

\begin{table*}[htbp]
\centering
\topcaption{
Fiducial definitions for the reported cross sections.
The common requirements are applied for both measurements.
}
\begin{tabular}{ll}
Cross section measurement & Fiducial requirements \\
\hline
Common requirements & $\pt^{\ell_1} > 20\GeV$ ,
  $\pt^{\ell_2} > 10\GeV$ ,
  $\pt^{\ell_{3,4}} > 5\GeV$ , \\
 & \\ [-1.9ex]
 & $\abs{\eta^{\ell}} < 2.5$,
  $m_{\ell^+ \ell^-} > 4\GeV$ (any opposite-sign same-flavor pair) \\
 & \\ [-1.7ex]
\hline
$\cPZ \to \elfour$ & $m_{\cPZ_1} > 40\GeV$ \\
 & $80 < m_{\elfour} < 100\GeV$ \\
\hline
$\cPZ\cPZ \to \elfour$   & $60 < m_{\cPZ_1}, m_{\cPZ_2} < 120\GeV$ \\
\end{tabular}
\label{table:fiducial_cuts}
\end{table*}

The measured cross sections are
\ifthenelse{\boolean{cms@external}}{
\begin{equation*}
\label{xs:z4l}
\begin{aligned}
  \sigma_{\text{fid}}&(\pp \to \cPZ \to \elfour) \\&= 30.5 ^{+5.2}_{-4.7}\stat  ^{+1.8}_{-1.4}\syst \pm 0.8\lum\unit{fb},\\
  \sigma_{\text{fid}}&(\pp \to \ZZ \to \elfour) \\&= 34.8 ^{+4.6}_{-4.2}\stat  ^{+1.2}_{-0.8}\syst \pm 0.9\lum\unit{fb}.
  \end{aligned}
\end{equation*}
}{
\begin{equation*}
\label{xs:z4l}
\begin{aligned}
  \sigma_{\text{fid}} (\pp \to \cPZ \to \elfour) &= 30.5 ^{+5.2}_{-4.7}\stat  ^{+1.8}_{-1.4}\syst \pm 0.8\lum\unit{fb},\\
  \sigma_{\text{fid}} (\pp \to \ZZ \to \elfour) &= 34.8 ^{+4.6}_{-4.2}\stat  ^{+1.2}_{-0.8}\syst \pm 0.9\lum\unit{fb}.
  \end{aligned}
\end{equation*}
}
The $\pp  \to  \cPZ  \to  \elfour$
fiducial cross section can be compared to
$27.9^{+1.0}_{-1.5} \pm 0.6\unit{fb}$ calculated at NLO in QCD with \POWHEG using
the same settings as used for the simulated sample described in Section~\ref{sec:mc}, with
dynamic scales $\mu_F = \mu_R = m_{\elfour}$.
The uncertainties are for scale and PDF variations, respectively.
The $\ZZ$ fiducial cross section can be compared to
$34.4^{+0.7}_{-0.6} \pm 0.5\unit{fb}$ calculated with \POWHEG and \MCFM
using the same settings as the simulated samples, with dynamic scales
$\mu_F = \mu_R = 0.5 m_{\elfour}$ for the
contribution from \MCFM.

The $\pp  \to  \cPZ  \to  \elfour$ fiducial cross section
is scaled to $\sigma (\pp \to \cPZ) \mathcal{B} (\cPZ \to 4\ell)$ using
the acceptance correction factor
$\mathcal{A} = 0.122 \pm 0.002$, estimated with \POWHEG.
This factor corrects the fiducial $\cPZ  \to  \elfour$ cross section to the
phase space with only the 80--100\GeV mass window and $m_{\ell^+\ell^-} > 4\GeV$
requirements, and also includes a correction, $0.96 \pm 0.01$,
for the contribution of nonresonant four-lepton production to
the signal region. The measured cross section is
\ifthenelse{\boolean{cms@external}}{
\begin{multline}
  \sigma (\pp \to \cPZ) \mathcal{B}(\cPZ \to \elfour) = \\
  250 ^{+43}_{-39}\stat  ^{+15}_{-11}\syst \pm 4\thy \pm 7\lum\unit{fb}.
\label{xstot:z4l}
\end{multline}
}{
\begin{equation}
  \sigma (\pp \to \cPZ) \mathcal{B}(\cPZ \to \elfour) = 250 ^{+43}_{-39}\stat  ^{+15}_{-11}\syst \pm 4\thy \pm 7\lum\unit{fb}.
\label{xstot:z4l}
\end{equation}
}
The branching fraction for the $\cPZ \to \elfour$ decay,
$\mathcal{B}(\cPZ \to \elfour)$,
is measured by comparing the cross section given by Eq.~(\ref{xstot:z4l})
with the $\cPZ \to \ell^+\ell^-$ cross section, and is computed as
\ifthenelse{\boolean{cms@external}}{
\begin{multline*}
\mathcal{B}(\cPZ \to \elfour) =\\
\frac{\sigma (\pp \to \cPZ \to \elfour)}
{\mathcal{C}^{\text{60--120}}_{\text{80--100}} \,
\sigma (\pp \to \cPZ \to \ell^+\ell^-) / \mathcal{B}(\cPZ \to \ell^+\ell^-)},
\end{multline*}
}{
\begin{equation*}
\mathcal{B}(\cPZ \to \elfour) =
\frac{\sigma (\pp \to \cPZ \to \elfour)}
{\mathcal{C}^{\text{60--120}}_{\text{80--100}} \,
\sigma (\pp \to \cPZ \to \ell^+\ell^-) / \mathcal{B}(\cPZ \to \ell^+\ell^-)},
\end{equation*}}
where $\sigma (\pp \to \cPZ \to \ell^+\ell^-) =
1870 _{-40}^{+50}\unit{pb} $ is the
$\cPZ  \to  \ell^+\ell^-$ cross section times branching fraction
calculated at NNLO with \textsc{fewz}~v2.0~\cite{Gavin:2010az} in the mass range
60--120\GeV. Its uncertainty includes PDF
uncertainties and uncertainties in $\alpha_s$, the charm and bottom quark masses,
and the effect of neglected higher-order corrections to the calculation.
The factor $\mathcal{C}^{\text{60--120}}_{\text{80--100}} = 0.926 \pm 0.001$ corrects for the
difference in $\cPZ$ mass windows and
is estimated using \POWHEG. Its uncertainty includes scale and PDF
variations.
The nominal $\cPZ$ to dilepton branching fraction
$\mathcal{B}(\cPZ \to \ell^+\ell^-)$
is 0.03366~\cite{Agashe:2014kda}. The measured value is
\ifthenelse{\boolean{cms@external}}{
\begin{multline*}
  \mathcal{B}(\cPZ \to \elfour) =\\ 4.9 _{-0.7}^{+0.8}\stat
  _{-0.2}^{+0.3}\syst _{-0.1}^{+0.2}\thy \pm 0.1\lum \times 10^{-6},
\end{multline*}
}{
\begin{equation*}
  \mathcal{B}(\cPZ \to \elfour) = 4.9 _{-0.7}^{+0.8}\stat
  _{-0.2}^{+0.3}\syst _{-0.1}^{+0.2}\thy \pm 0.1\lum \times 10^{-6},
\end{equation*}
}
where the theoretical uncertainty includes the uncertainties in $\mathcal{A}$,
$\mathcal{C}^{\text{60--120}}_{\text{80--100}}$, and
$\sigma (\pp \to \cPZ) \mathcal{B} (\cPZ \to \ell^+\ell^-)$.
This can be compared with $4.6 \times 10^{-6}$, computed with
\textsc{MadGraph5\_aMC@NLO}~\cite{mg_amcnlo},
and is consistent with the CMS and ATLAS measurements at
$\sqrt{s} = 7$ and 8\TeV~\cite{CMS:2012bw, Aad:2014wra}.

The total $\ZZ$ production cross section for both dileptons produced in the
mass range 60--120\GeV and $m_{\ell^+\ell^{\prime -}} > 4\GeV$ is found to be
\ifthenelse{\boolean{cms@external}}{
\begin{multline*}
 \sigma(\pp \to \ZZ) =\\ 14.6 ^{+1.9}_{-1.8}
\stat  ^{+0.5}_{-0.3}\syst \pm 0.2\thy  \pm 0.4\lum\unit{pb}.
\end{multline*}
}{
\begin{equation*}
  \sigma(\pp \to \ZZ) = 14.6 ^{+1.9}_{-1.8}
\stat  ^{+0.5}_{-0.3}\syst \pm 0.2\thy  \pm 0.4\lum\unit{pb}.
\end{equation*}
}
The measured total cross section can be compared to the theoretical value of
$14.5^{+0.5}_{-0.4} \pm 0.2\unit{pb}$ calculated
with a combination of \POWHEG and \MCFM with the same settings as described for
$\sigma_{\text{fid}} (\pp \to \ZZ \to \elfour)$.
It can also be compared to
$16.2^{+0.6}_{-0.4}$\unit{pb}, calculated at NNLO in QCD via
\textsc{matrix}~\cite{Cascioli:2014yka,Grazzini:2015hta}, or
$15.0^{+0.7}_{-0.6} \pm 0.2$\unit{pb}, calculated with \MCFM at NLO in QCD with
additional contributions from LO $\Pg\Pg  \to  \cPZ\cPZ$ diagrams. Both values are
calculated with the NNPDF3.0 PDF sets, at NNLO and NLO respectively, and fixed
scales set to $\mu_F = \mu_R = m_\cPZ$.

\begin{figure}[htbp]
\centering
\includegraphics[width=\cmsFigWidth]{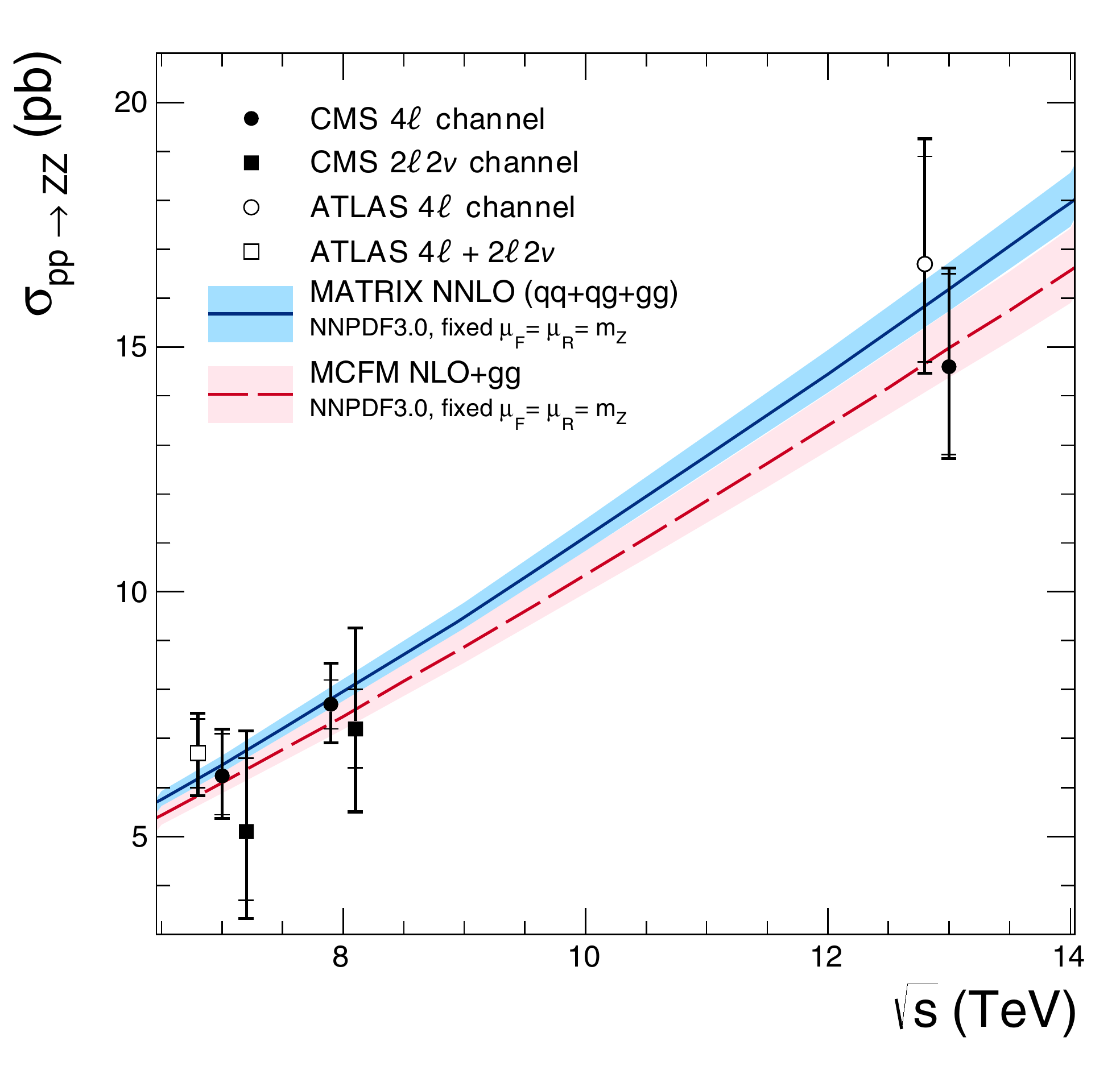}
\caption{
  The total $\ZZ$ cross section as a function of the proton-proton
  center-of-mass energy. Results from the CMS and ATLAS experiments are
  compared to predictions from \textsc{matrix} and \MCFM with
  NNPDF3.0 PDF sets and fixed scales $\mu_F = \mu_R = m_\cPZ$. Details of
  the calculations and uncertainties are given in the text.
  Measurements at the same center-of-mass energy are shifted slightly along
  the x-axis for clarity.
}
\label{fig:xsec_vs_sqrts}
\end{figure}

The total $\ZZ$ cross section is shown
in Fig.~\ref{fig:xsec_vs_sqrts} as a function of the proton-proton
center-of-mass energy. Results from the
CMS~\cite{Chatrchyan:2012sga, CMS:2014xja, Khachatryan:2015pba} and
ATLAS~\cite{Aad:2012awa, Aad:2015rka, Aad:2015zqe} experiments are
compared to predictions from \textsc{matrix} and \MCFM with
the NNPDF3.0 PDF sets and fixed scales $\mu_F = \mu_R = m_\cPZ$.
The \textsc{matrix} prediction uses PDFs calculated at NNLO, while the
\MCFM prediction uses NLO PDFs.
The uncertainties are statistical (inner bars) and
statistical and systematic added in quadrature (outer bars). The band
around the \textsc{matrix} predictions reflects scale uncertainties, while
the band around the \MCFM predictions reflects both scale and PDF
uncertainties. The theoretical predictions and all CMS measurements are
performed in the dilepton mass range 60--120\GeV. All ATLAS measurements
are in the mass window 66--116\GeV. The smaller mass window is
estimated to cause a 1.6\% reduction in the measured cross section.

\section{Summary}

Results have been presented for a study of four-lepton final states in
proton-proton collisions at $\sqrt{s} = 13\TeV$ with the CMS detector at the
LHC. The $\pp \to \ZZ$ cross section has been measured to be
$\sigma(\pp \to \ZZ) = 14.6 ^{+1.9}_{-1.8}\stat ^{+0.5}_{-0.3}\syst
\pm 0.2\thy \pm 0.4\lum\unit{pb}$ for $\cPZ$ boson masses in the range
$60 < m_{\cPZ} < 120\GeV$. The branching fraction for
$\cPZ$ boson decays to four leptons has been measured to be
$\mathcal{B}(\cPZ \to \elfour) = 4.9 _{-0.7}^{+0.8}\stat
_{-0.2}^{+0.3}\syst _{-0.1}^{+0.2}\thy \pm 0.1\lum \times 10^{-6}$
for four-lepton mass in the range
$80 < m_{\elfour} < 100\GeV$ and dilepton mass
$m_{\ell^+\ell^-} > 4\GeV$ for all oppositely charged same-flavor lepton pairs. The results
are consistent with SM predictions.

\begin{acknowledgments}
\hyphenation{Rachada-pisek}
We thank Massimiliano Grazzini and his collaborators for providing the NNLO
cross section calculations.
We congratulate our colleagues in the CERN accelerator departments for the excellent performance of the LHC and thank the technical and administrative staffs at CERN and at other CMS institutes for their contributions to the success of the CMS effort. In addition, we gratefully acknowledge the computing centers and personnel of the Worldwide LHC Computing Grid for delivering so effectively the computing infrastructure essential to our analyses. Finally, we acknowledge the enduring support for the construction and operation of the LHC and the CMS detector provided by the following funding agencies: BMWFW and FWF (Austria); FNRS and FWO (Belgium); CNPq, CAPES, FAPERJ, and FAPESP (Brazil); MES (Bulgaria); CERN; CAS, MoST, and NSFC (China); COLCIENCIAS (Colombia); MSES and CSF (Croatia); RPF (Cyprus); SENESCYT (Ecuador); MoER, ERC IUT and ERDF (Estonia); Academy of Finland, MEC, and HIP (Finland); CEA and CNRS/IN2P3 (France); BMBF, DFG, and HGF (Germany); GSRT (Greece); OTKA and NIH (Hungary); DAE and DST (India); IPM (Iran); SFI (Ireland); INFN (Italy); MSIP and NRF (Republic of Korea); LAS (Lithuania); MOE and UM (Malaysia); BUAP, CINVESTAV, CONACYT, LNS, SEP, and UASLP-FAI (Mexico); MBIE (New Zealand); PAEC (Pakistan); MSHE and NSC (Poland); FCT (Portugal); JINR (Dubna); MON, RosAtom, RAS and RFBR (Russia); MESTD (Serbia); SEIDI and CPAN (Spain); Swiss Funding Agencies (Switzerland); MST (Taipei); ThEPCenter, IPST, STAR and NSTDA (Thailand); TUBITAK and TAEK (Turkey); NASU and SFFR (Ukraine); STFC (United Kingdom); DOE and NSF (USA).

Individuals have received support from the Marie-Curie program and the European Research Council and EPLANET (European Union); the Leventis Foundation; the A. P. Sloan Foundation; the Alexander von Humboldt Foundation; the Belgian Federal Science Policy Office; the Fonds pour la Formation \`a la Recherche dans l'Industrie et dans l'Agriculture (FRIA-Belgium); the Agentschap voor Innovatie door Wetenschap en Technologie (IWT-Belgium); the Ministry of Education, Youth and Sports (MEYS) of the Czech Republic; the Council of Science and Industrial Research, India; the HOMING PLUS program of the Foundation for Polish Science, cofinanced from European Union, Regional Development Fund, the Mobility Plus program of the Ministry of Science and Higher Education, the National Science Center (Poland), contracts Harmonia 2014/14/M/ST2/00428, Opus 2013/11/B/ST2/04202, 2014/13/B/ST2/02543 and 2014/15/B/ST2/03998, Sonata-bis 2012/07/E/ST2/01406; the Thalis and Aristeia programs cofinanced by EU-ESF and the Greek NSRF; the National Priorities Research Program by Qatar National Research Fund; the Programa Clar\'in-COFUND del Principado de Asturias; the Rachadapisek Sompot Fund for Postdoctoral Fellowship, Chulalongkorn University and the Chulalongkorn Academic into Its 2nd Century Project Advancement Project (Thailand); and the Welch Foundation, contract C-1845.
\end{acknowledgments}

\bibliography{auto_generated}

\cleardoublepage \appendix\section{The CMS Collaboration \label{app:collab}}\begin{sloppypar}\hyphenpenalty=5000\widowpenalty=500\clubpenalty=5000\textbf{Yerevan Physics Institute,  Yerevan,  Armenia}\\*[0pt]
V.~Khachatryan, A.M.~Sirunyan, A.~Tumasyan
\vskip\cmsinstskip
\textbf{Institut f\"{u}r Hochenergiephysik der OeAW,  Wien,  Austria}\\*[0pt]
W.~Adam, E.~Asilar, T.~Bergauer, J.~Brandstetter, E.~Brondolin, M.~Dragicevic, J.~Er\"{o}, M.~Flechl, M.~Friedl, R.~Fr\"{u}hwirth\cmsAuthorMark{1}, V.M.~Ghete, C.~Hartl, N.~H\"{o}rmann, J.~Hrubec, M.~Jeitler\cmsAuthorMark{1}, A.~K\"{o}nig, I.~Kr\"{a}tschmer, D.~Liko, T.~Matsushita, I.~Mikulec, D.~Rabady, N.~Rad, B.~Rahbaran, H.~Rohringer, J.~Schieck\cmsAuthorMark{1}, J.~Strauss, W.~Treberer-Treberspurg, W.~Waltenberger, C.-E.~Wulz\cmsAuthorMark{1}
% ...
\vskip\cmsinstskip
% ...
1:~~Also at Vienna University of Technology, Vienna, Austria\\
% ...

\end{sloppypar}
\end{document}